\documentclass{mn2e}
\usepackage{epsfig}
\usepackage{times}
\begin{document}

\title[H$\alpha$ in X-ray binaries] {An anti-correlation between X-ray luminosity and H$\alpha$
equivalent width in X-ray binaries} 
\author[Fender et al.]  {R. P. Fender$^{1,2}$,
D.M. Russell$^{1,2}$, C. Knigge$^1$, R.Soria$^3$, R.I. Hynes$^4$, 
M.Goad$^5$\\ $^1$ School of Physics and Astronomy, University of
Southampton, Southampton SO17 1BJ\\ $^2$Astronomical Institute `Anton
Pannekoek', University of Amsterdam, Kruislaan 403, 1098 SJ, Amsterdam, The Netherlands\\ 
$^3$ Mullard Space Science Laboratory (UCL), Holmbury St Mary, Dorking, Surrey RH5 6NT\\
$^4$ Department of Physics and Astronomy, Louisiana State University, Baton Rouge, LA 70803\\ $^5$ Department of Physics and Astronomy, University of Leicester, LE1 7RH\\} \maketitle

\begin{abstract}

We report an anticorrelation between continuum luminosity and the
equivalent width (EW) of the H$\alpha$ emission line in X-ray binary
systems. The effect is evident both in a universal monotonic increase
in H$\alpha$ EW with time following outbursts, as systems fade, and in
a comparison between measured EWs and contemporaneous X-ray
measurements. The effect is most clear for black hole binaries in the
low/hard X-ray state, which is prevalent at X-ray luminosities below
$\sim 1$\% Eddington. We do not find strong evidence for significant
changes in line profiles across accretion state changes, but this is
hampered by a lack of good data at such times.  The observed
anti-correlation, highly significant for black hole binaries, is only
marginally so for neutron star systems, for which there are far less
data. Comparison with previously established correlations between
optical and X-ray luminosity suggest that the line luminosity is
falling as the X-ray and optical luminosities drop, but not as fast
(approximately as $L_{\rm H\alpha} \propto L_{\rm X}^{\sim 0.4}
\propto L_{\rm opt}^{\sim 0.7}$).  We briefly discuss possible origins for
such an effect, including the optical depth, form of the irradiating
spectrum and geometry of the accetion flow. Further refinement of the
relation in the future may allow measurements of H$\alpha$ EW to be
used to estimate the luminosity of, and hence the distance to, X-ray
binary systems. Beyond this, further progress will require a better
sample of spectro-photometric data.

\end{abstract}


\section{Introduction}

The process of accretion is the power source driving the luminosities
for a wide range of objects, including protostars, binary systems
containing accreting white dwarfs, neutron stars or black holes, gamma
ray bursts and supermassive black holes in active galactic nuclei. A
comprehensive review of this process is provided by Frank, King \&
Raine (2002).

In most of these systems, most of the time, the accretion process
proceeds via an accretion disc which transports angular momentum
outwards and matter inwards. The temperature of this disc increases
towards the centre, and is a function of accretion rate and central
accretor mass. The disc may also produce (or even be partially
replaced by) at various times a relatively cool disc wind, a very hot
corona and a collimated relativistic outflow or `jet'. Finally, the
`state' of the accretion flow at the centre of the disc may vary
dramatically indicating rapid variations between phases with different
geometries, temperatures and outflows. Reviews of observation of
accretion onto white dwarfs, neutron stars and black holes in binary
systems can be found in e.g. Warner (2003) and Lewin \& van der Klis
(2006).

In X-ray binary systems containing an accreting neutron star or black
hole the region of the accretion disc responsible for the optical
continuum and emission lines lies at a large distance from the central
accretor (several light seconds, or $\geq 10^4$ gravitational radii --
e.g. Hynes et al. 2006).  Many such systems are transients, in that
they display phases of very high luminosities lasting typically weeks
to months, followed by long periods in quiescence (e.g. Chen, Shrader
\& Livio 1997). Such cycles are likely to have their origin in
accretion disc instabilities associated with hydrogen ionization
(Frank et al. 2002 and references therein).

It is widely accepted that the emission lines in such systems arise in
the rotating accretion disc flow. The strongest evidence for this
interpretation is in the form of twin-peaked line profiles (e.g. Horne
\& Marsh 1986; Charles \& Coe 2006 and references therein).  The
dominant pumping mechanism for these lines is likely to be 
irradiation by the central X-ray source at high luminosities, as is
observed for the optical continuum (van Paradijs \& McClintock 1994). At lower
luminosities viscous heating of the disc may contribute significantly.

However, the picture may not be so simple. Russell et al. (2006 and
references therein) have demonstrated that in black hole X-ray
binaries (BHXBs) in `hard' X-ray states synchrotron emission from the
jet dominates the continuum in the near-infrared and can contribute
significantly in the optical band. Wu et al. (2001, 2002) argue for
three distinct origins for the emission line depending on the source
luminosity and spectral states. In their model, in bright/soft X-ray
states the line arises in the atmosphere of an optically thick disc,
whereas in bright/hard X-ray states the H$\alpha$ line arises in a
dense outflow.  In the faintest quiescent states the whole accretion flow
is optically thin. 

In a related work, Eikenberry et al. (1998) found a near-linear
correlation between Br$\gamma$ (2.166 $\mu$m) integrated line flux and
the adjacent continuum flux density in the black hole X-ray binary GRS
1915+105, during phases when the continuum was likely to be
synchrotron emission from the powerful jet in this source. This
suggests radiative pumping of the lines by UV emission from the jet.
Comparison with phases of soft X-ray spectra with no jet emission in
the same study were taken to indicate that the jet is a far better
source of photoionising photons than the inner accretion disc,
presumably due to the fact that the ultraviolet-emitting region of the
jet is raised above the disc and illuminates it very efficiently (this
argument does not work if this region of the jet is moving
relativistically away from the disc and therefore strongly Doppler
de-boosted). However, a strong change in H$\alpha$ EW associated with
X-ray state, and therefore jet production, is {\em not} observed in GX
339-4 (Wu et al. 2001; see also Fig 1), possibly arguing against
strong pumping of H$\alpha$ by the jet.

In this paper we have compiled measurements of the strength of
H$\alpha$, the most prominent emission line in the optical spectra of
X-ray binaries. Our goal was to see how the properties of the emission
line varied, if at all, with accretion `state' and luminosity of X-ray
binary systems, and whether or not there was clear evidence for phases
where the lines were formed in an outflow rather than a disc.  What we
find is evidence for an anti-correlation between the equivalent width
(ratio of line to local continuum flux) and overall luminosity of such
systems, in particular when in the `hard' X-ray state (at luminosities
below about 1\% of the Eddington limit).

\begin{table*}
\begin{tabular}{cccc}
Source & References \\
\hline
Black hole binaries & \\
\hline
GRO J0422+32 & Callanan et al. 1995, Bonnet-Bidaud \& Mouchet 1995, Garcia et al. 1996, Casares et al. 1995, Harlaftis et al. 1999\\
A 0620-00 & Whelan et al. 1977, Murdin et al. 1980, Marsh, Robinson \& Wood 1994 \\
XTE J1118+480 & Torres et al. 2004, Zurita (private communication) \\
GS 1124-68 &  Orosz et al. 1994, della Valle et al. 1998, Casares et al. 1997, Sutaria et al. 2002, Ebisawa et al. 1994 \\
XTE J1550-564 & Orosz et al. 2002, Sanchez-Fernandez et al. 1999, Casares, Dubus \& Homer 1999, Buxton et al. 1999 \\
GRO J1655-40 & Bianchini et al. 1997, Shahbaz et al. 2000, Soria et al. 2000 \\
GX 339-4 & Grindlay 1979, Soria, Wu \& Johnston 1999, Shahbaz, Fender \& Charles 2001 \\
GS 2000+25 & Charles et al. 1988, Shahbaz et al. 1996, Tsunemi et al. 1989, Casares et al. 1995, Garcia et al. 2001 \\
XTE J1650 & Augusteijn et al. 2001, Sanchez-Fernandez et al. 2002 \\
XTE J1859+226 & Garnavich et al. 1999, Wagner et al. 1999, Zurita et al. 2002 \\
GRS 1009-45 & della Valle et al. 1997 \\
V404 Cyg & Casares et al. 1991, 1993, Hynes et al. (2004) \\
4U 1957+11 & Shahbaz et al. 1996 \\
\hline
Neutron star binaries & \\
\hline
Aql X-1 & Charles et al. 1980, Shahbaz et al. 1996, Garcia et al. 1999, Shahbaz et al. 1998 \\
Cen X-4 & van Paradijs et al. 1980 and 1987, Shahbaz et al. 1996, Campana et al. 2004 \\
Sco X-1 & Shahbaz et al. 1996 \\
Cyg X-2 & Shahbaz et al. 1996 \\
EXO 0748-676 & Pearson et al. 2006 \\
IGR 00291+5934 & Torres et al. 2004, Filippenko et al. 2004 \\
SAX J1808.4-3658 & Campana et al. 2004, Campana \& Stella 2004 \\
XTE J1814-338 & Steeghs 2003, Papitto et al. 2007 \\
HETE J1900.1-2455 & Elebert et al. 2008 \\
XTE J2123-058 & Casares et al. 2002, Tomsick et al. 2004 \\
\hline
\end{tabular}
\caption{X-ray binaries used in this study, with relevant references.}
\end{table*}

\section{An anti-correlation between H$\alpha$ EW and luminosity in black hole X-ray binaries}

\begin{figure}
\centerline{\epsfig{file=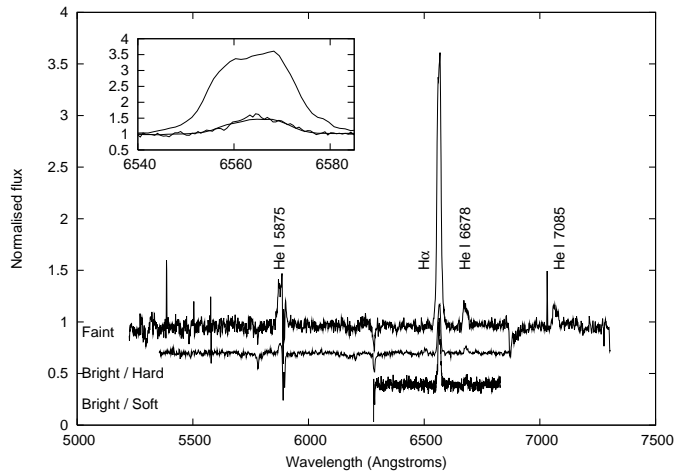, angle=270, width=9cm}}
\caption{Normalised optical spectra of GX 339-4 in low luminosity,
bright/hard (-0.3 in normalisation) and bright/soft (-0.6) states
(Soria, Wu \& Johnston 1999; Shahbaz, Fender \& Charles 2001).  These
spectra are almost certainly uncontaminated by a companion star. The
strongest emission line is H$\alpha$, which clearly has a much greater
equivalent width in the faint state than in either of the brighter
states. The inset shows the H$\alpha$ profile for all three spectra
plotted on the same scale. The bright/hard and bright/soft spectra,
with H$\alpha$ EW $\sim -7$\AA\ were simultaneous with photometric
observations which recorded $V \sim 16.5$; the faint state
observations, with H$\alpha \sim -55$ \AA\ were obtained when $R = 20.1
\pm 0.1$.}
\end{figure}

Two approaches, discussed below, are taken in order to investigate the
relation between emission line strength and luminosity in black hole
X-ray binaries. In the first we track the equivalent width (EW) of
H$\alpha$ as a function of time following an X-ray outburst, in which
case the general trend of the relation with luminosity is inferred
(albeit quite confidently); in the second approach we directly compare
H$\alpha$ EW with X-ray luminosity, a more direct approach but one
which is limited by a lack of data at low luminosities.

Before doing so, we present in Fig 1 optical spectra in the H$\alpha$
region for the black hole X-ray binary GX 339-4 in three `states': a
very faint state (upper spectrum), a bright state with a hard X-ray
spectrum (middle spectrum) and a bright state with a soft X-ray
spectrum (lower spectrum). What is immediately clear is that in the
faint state the H$\alpha$ EW is much greater than it is in either of
the higher-luminosity states, and also that there does not appear to
be much change in the line EW (or profile) between the two
high-luminosity states. Indications of the optical magnitudes and EWs
at the different epochs are given in the figure caption.  See Soria,
Wu \& Johnston (1999) and Shahbaz, Fender \& Charles (2001) for a
detailed discussion of these spectra.

\subsection{Trends of increasing H$\alpha$ EW during outburst decays}

\begin{figure}
\centerline{\epsfig{file=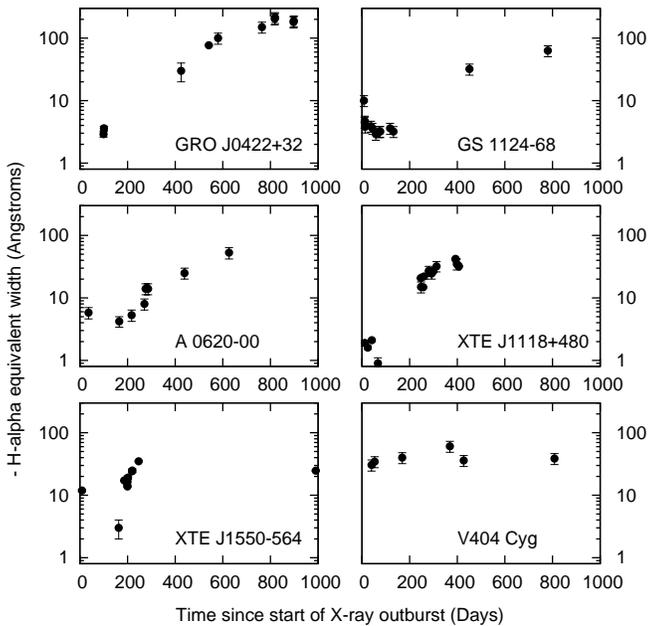, angle=270, width=12.5cm}}
\caption{Variation of H$\alpha$ equivalent width (EW) with time since
X-ray discovery of six transient black hole binaries. For the top four
sources, GRO J0422+32, GS 1124-68, A0620-00, XTE J1118+480, there is a
clear increase in the EW with time. For XTE J1550-564 there is an
initial increase but hints of a subsequent decline, and for V404 Cyg
there is no clear trend. Assuming this is a representative sample,
this strongly suggests an anti-correlation of EW with luminosity as the
majority of black hole transients fade following outbursts.}
\end{figure}

\begin{figure}
\centerline{\epsfig{file=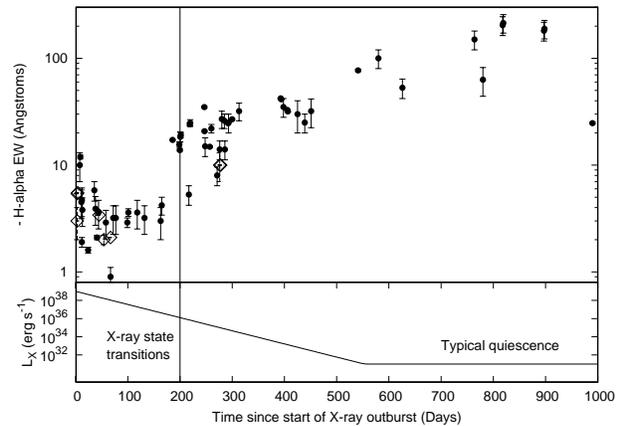, angle=270, width=9.5cm}}
\caption{The data for the first five sources in Fig 1 overplotted
(filled circles), plus a small number of additional measurements (open
diamonds; see text for details). The similarity in the post-outburst
time evolution of the H$\alpha$ EW is remarkable. In the lower panel
is a crude approximation of the X-ray luminosity evolution of the source
based upon the typical exponential decay timescale for transients
(Chen, Shrader \& Livio 1997) and typical quiescent level (Remillard
\& McClintock 2006 and references therein). The vertical line at 200
days delay indicates the time after which all the sources should be in
the hard spectral state; at earlier times they could be in either soft
or hard X-ray states (Maccarone 2003; Homan \& Belloni 2005).}
\end{figure}

Most X-ray transients (with both black hole and neutron star
accretors) follow a monotonic decay (with some occasional
rebrightenings) in X-ray luminosity after the first few weeks or
months of outburst. Specifically, Chen, Shrader \& Livio (1997) noted
a mean exponential decay timescale for X-ray transients of around 30
days in the initial decline from outburst. Such a decay rate indicates
that within 2 years a source should have returned to a 'quiescent'
level, typically a factor $\sim 10^7$ fainter in $L_X$ than at the
peak of outburst. As a caveat it should be noted that the observed
decays are often far from simple exponentials. Nevertheless, a trend
of increasing EW with time following outburst would therefore be a
strong indicator that there is a luminosity -- EW anti-correlation.
This effect is investigated in Fig 2, where time evolution of
H$\alpha$ equivalent width is plotted for the first 1000 days
following outburst, for six BHXBs, which for most sources should
include the return to quiescence. The dates and references for these
outbursts are given in Table 1.  These were the only six sources for
which we could find good coverage of the H$\alpha$ EW following the
outburst. A clear trend of increasing EW with time is observed in the
first four sources (GRO J0422+32, GS 1124-68, A0620-00, XTE
J1118+480), following an intial period of erratic variations or dips,
which probably corresponds to points before the monotonic decay had
begun. The fifth source, XTE J1550-564 shows a similar pattern of
behaviour, but there are hints, at around 1000 days, of a subsequent
decline in the EW. Inspection of the RXTE ASM monitoring at this time
does not reveal any obvious subsequent outburst, but is not sensitive
to activity below $L_X \sim 10^{36}$ erg s$^{-1}$. However, optical
monitoring reported by Orosz et al. (2002) does in fact indicate that
this final EW measurement was within 200 days of an additional optical
outburst; therefore its behaviour may not be discrepant. Note that for
some of the other sources there is evidence in the literature for
minor rebrightenings after 200 days which should not affect the broad
conclusions here but may have some relevance for the poorer
anticorrelation between X-ray luminosity and quasi-simultaneous EW
measurements in the next section. Finally in the sixth panel V404 Cyg,
which was in several ways an unusual transient (e.g. Kitamoto et
al. 1989) and is far more luminous in `quiescence' than most BHXBs,
does not display any clear trend of increasing EW following its
outburst.

Therefore, it does seem that for the majority (5 of 6) of the sources,
there is an increase in H$\alpha$ EW following outburst, which
strongly suggests an anti-correlation between luminosity and H$\alpha$
EW in these systems. We have no reason to believe that our sample is
strongly biased, but the sample of available EW data is clearly small
and it is probably too early to quantify the diversity in these
general EW trends.

In Fig 3 we overplot the data for the first five sources from Fig 2;
i.e. all except V404 Cyg.  The Spearman rank correlation coefficient
for the increase in EW with time for the sample of all six sources is
$r_s = 0.82$, a rank correlation at the $6.8\sigma$ level. Removing
V404 Cyg, the most discrepant source and arguably a `non-standard'
transient (i.e. the data set plotted in Fig 3), increases these
figures slightly to $r_s = 0.89$, corresponding to a rank correlation
at the $7.1\sigma$ level. Adding a small number of additional points
for the sources GS 2000+25, XTE J1650-500, XTE J1859+226 and GRS
1009-45 increases the significance of the correlation to $7.3\sigma$
(see Table 1 for references).  Table 2 presents a compilation of the
Spearman rank correlation coefficients and their significance for 
most of the samples discussed in this paper.

Fig 3 does however also hint at a complication to the picture, namely
that the correlation between elapsed time and H$\alpha$ EW is clearer
after about 200 days, which is around the time that most sources are
expected to have made a transition back to the hard X-ray spectral
state, at a (2--10 keV) X-ray luminosity of a few $\times 10^{37}$ erg
s$^{-1}$ (Maccarone 2002; Homan \& Belloni 2005). At this point the
steady, powerful jet associated with this state is expected to be
reactivated after a period in the soft X-ray state in which it was
suppressed (Fender, Belloni \& Gallo 2004 and references therein). In
order to investigate this effect we separated the data from Fig 3 at
the point of 200 days (vertical line in Fig 3). The data after this
point, almost certainly exclusively in the hard X-ray state, still
showed a significant correlation between elapsed time and the
H$\alpha$ EW at the 5.1$\sigma$ level ($r_s = 0.80$). The sample of
data prior to 200 days were not significantly correlated in any way
with the elapsed time. The increase in overall significance for the
entire sample can be attributed to the combination of the correlation
in the hard X-ray state together with a generally lower EW in the
`clump' of data at shorter elapsed times, which will include a mix of
luminous hard and soft X-ray states. Fig 3 is a log-linear plot,
therefore it is clear from the apparent linear relation that an
exponential fit should be appropriate. A fit to the ensemble of data
from 200 days onwards (excluding V404 Cyg), i.e. almost certainly in
the hard state, indicates an e-folding time constant for the increase
of H$\alpha$ EW of $286 \pm 23$ days. If the 30 days e-folding decay
time for the X-ray luminosity were an accurate estimate, then this
would imply that EW $\propto L_X^{-0.1}$ (approximately). We will test
this in the next section. Note that the global correlation between
X-ray and optical luminosities for BHXBs reported in Russell et
al. (2006; see also van Paradijs \& McClintock 1994) demonstrates that
any anti-correlation found, or inferred, between EW and X-ray
luminosity, is also one between EW and optical luminosity (hence the
use of the generic term `luminosity' in much of what follows).

\begin{table}
\begin{tabular}{ccc}
\hline
Source & EW vs. time & EW vs. $L_X$ \\
\hline
GRO J0422+32 & 0.90 (2.7$\sigma$) & -0.90 (1.8$\sigma$) \\
GS 1124-68 & -0.10 (0.3$\sigma$) & \\
A 0620-00 & 0.92 (2.4$\sigma$) & -0.56 (1.1$\sigma$) \\
XTE J1118+480 & 0.94 (4.2$\sigma$) &  \\
XTE J1550-564 & 0.93 (2.9$\sigma$) & -0.91 (2.9$\sigma$) \\
V404 Cyg & 0.54 (1.2$\sigma$) & \\
GX 339-4 & & -0.49 (1.1$\sigma$) \\
GRO J1655-40 & & 0.41 (1.7$\sigma$) \\
\hline
BH Ensemble & 0.82 (6.8$\sigma$) & -0.48 (3.9$\sigma$) \\ 
\hline
NS Ensemble & -- & -0.68 (2.3$\sigma$) \\
\hline
BH+NS Ensemble & -- & -0.58 (5.1$\sigma$) \\
\hline
\end{tabular}
\caption{Spearman rank correlation coefficients, and their
  significance in terms of standard deviations, for the relation of
  H$\alpha$ equivalent width as a function of time since outburst
  (column 2) and X-ray luminosity (column 3). The ensemble values are
  presented in the last row. Gaps indicate sources for which there are
  too little data to make meaningful tests.}
\end{table}

\subsection{EW as a function of X-ray luminosity}

\begin{figure}
\centerline{\epsfig{file=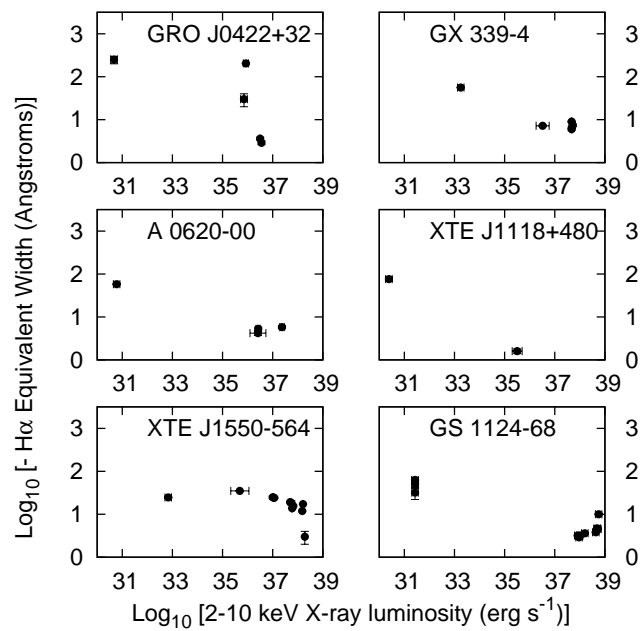, angle=270, width=12.5cm}}
\caption{Variation of H$\alpha$ EW with estimated X-ray luminosity
based on contemporaneous observations. The sources are the same as
presented in Fig 2, except that V404 Cyg has been replaced with GX
339-4. All sources show a general anti-correlation, although the
patterns of behaviour are clearly not identical.}
\end{figure}

\begin{figure}
\centerline{\epsfig{file=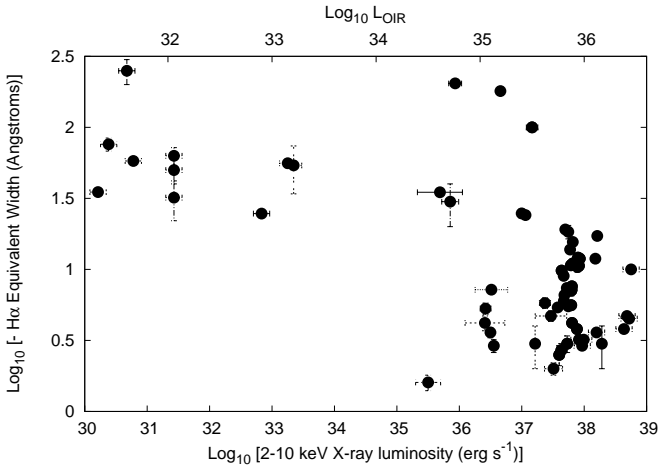, angle=270, width=9cm}}
\caption{ H$\alpha$ equivalent width (EW) as a function of
contemporaneously measured X-ray luminosity (typically 2--10 keV). All
the data from Fig 4 are combined, as well as a small number of
additional measurements (see text for details). There is an overall
rank anti-correlation, at a $3.9\sigma$ confidence level. The upper
x-axis indicates an estimate of the optical--infrared luminosity,
$L_{\rm OIR}$, calculated from the relation $L_{\rm OIR} = 10^{13.2}
L_{\rm X}^{0.60}$ for black hole binaries presented in Russell et
al. (2006).}
\end{figure}

The approach taken in section 2.1, in plotting H$\alpha$ EW as a
function of time for BHXBs is a strong but indirect indication of an
anti-correlation of the EW with luminosity. A more direct comparison of
luminosity and EW would be desirable, to rule out some unexpected
mechanism rather than luminosity changes dominating the observed
effect.  Unfortunately, the majority of the optical spectroscopic observations
do not appear to be well flux-calibrated, so we cannot directly
compare EW with optical continuum luminosity. 

A second possibility is to compare with an X-ray measurement which has
been made contemporaneously, which is what we attempt next. In order
to be considered, the X-ray and optical spectroscopic measurements had
to be made within one day of each other, except at quiescent levels at
which we assumed a more or less stable level had been reached
(quiescent variability can reach factors of several, but this not very
significant for these logarithmic plots). In Fig 4 we plot H$\alpha$
EW as a function of $L_X$ for the sample of objects discussed
previously (except for V404 Cyg for which we do not have
contemporaneous X-ray measurements), plus the more unusual object GX
339-4. None of the individual sources show a statistically significant
rank (anti-)correlation (see Table 2). In Fig 5 we plot the ensemble
of data points, plus a small amount of additional data from the
sources V404 Cyg, GRO J1655-40, 4U 1957+11, GS 2000+25, XTE J1650-500
and XTE J1859+226 (see Table 1 for references).

A Spearman rank correlation test for the complete sample plotted in
Fig 5 results in a Spearman rank correlation coefficient of $r_s =
-0.48$ (a $\sim 3.9\sigma$ result). This supports the
conclusion of section 2.1. A single power-law fit to the data
presented in Fig 5 gives

\[
EW = (-24 \pm 18) \frac{L_X}{10^{36}}^{-0.18 \pm 0.06}
\] 

This is close to, but steeper than, the EW $\propto L_X^{-0.1}$
estimated in section 2.1, and indicates a mean X-ray decay e-folding
time less than the 30 days given in Chen et al. (1997). The apparent
anti-correlation seems to be dominated by the difference between
measurements at $L_X \leq 10^{36}$ erg s$^{-1}$, where all sources are
in the hard X-ray spectral state, and those at higher luminosities,
where transitions between different spectral states can occur (see
e.g. Nowak 1995; Homan \& Belloni 2005; Remillard \& McClintock 2006
for a discussion of these states).

It is somewhat puzzling that the anti-correlation stands out much more
clearly in the first analysis (Figs 2 and 3) than in the second (Figs
4 and 5), which implies that there may not be a simple one-to-one
relation between X-ray luminosity and H$\alpha$ EW. In this context it
is interesting to note that the low-frequency QPOs in some hard state
black hole candidates show similar monotonic behaviour with time which
is not so simple when compared to X-ray flux (e.g. XTE J1118+480 in
Wood et al. 2001). In any case, the anti-correlation, whether simple or
not, is clearly there.

\subsubsection{Neutron star X-ray binaries}

Far fewer results were available in the literature for neutron star
systems; the details and references are provided in Table 1.  In Fig 6
we plot the neutron star data alongside the black hole data presented
in Fig 5. What we see is approximately similar behaviour, in that
there is a suggestion of a similar trend, with a similar (or slightly
lower) normalisation. In fact a Spearman rank correlation test
indicates that the sample of seven neutron star measurements are
anticorrelated at the $2.7 \sigma$ level ($r_s = -0.84$). So the
neutron star data are consistent with, but do not independently
establish, the anti-correlation found for the BHXBs.

\begin{figure}
\centerline{\epsfig{file=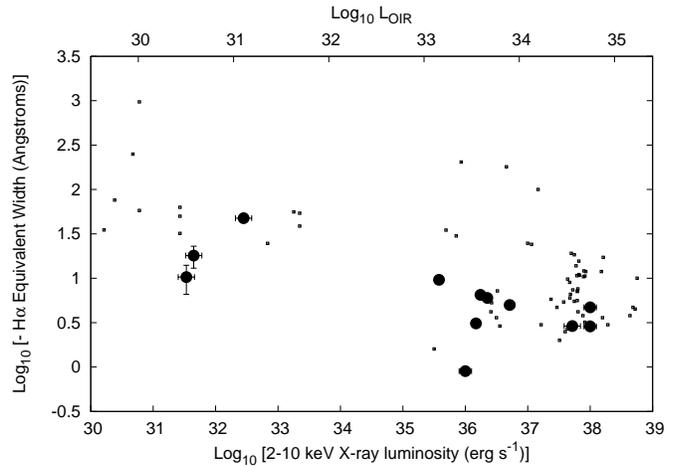, angle=270, width=9cm}}
\caption{As Fig 5, but for neutron star binaries (the black hole
sample from Fig 5 are indicated as small dots).  These points are
consistent with neutron star X-ray binaries following a similar
anti-correlation to black hole binaries (with a hint of a lower
normalisation). As in Fig 4, the upper x-axis indicates an estimate of
$L_{\rm OIR}$ based upon the relation from Russell et al. (2006),
which is $L_{\rm OIR} = 10^{9.7}L_{\rm X}^{0.66}$, slightly different
to that for black holes (hence the slightly different range in $L_{\rm
OIR}$).  }
\end{figure}

Adding the BHXB and neutron star samples results in a Spearman rank
correlation coefficient of $r_s = -0.58$, corresponding to an
anticorrelation at the $5.1 \sigma$ level. 

In summary, although the individual simultaneous $L_X$ and EW
measurements do not confirm the anti-correlation further, the effect
is significant when considering the ensemble of all data points,
moreso when combining neutron star data with the black hole sample.

\subsection{Relation to optical continuum luminosity}

The upper x-axes of Figs 5 and 6 also indicate a rough estimate of the
optical--infrared luminosities, $L_{\rm OIR} = \nu L_{\nu}$ based on
the relations presented in Russell et al. (2006), and repeated in the
figure captions (without statistical uncertainties). Note that the
relations are slightly different for black holes and neutron stars
($L_{\rm OIR}$ is larger for black holes at any given X-ray
luminosity). This allows us to get a better idea of how the EW varies
as a function of the optical continuum.

Subject to caveats about the possibly complex nature of the relation
to the lines, we can go further. By definition, 

\[
{\rm EW} \propto \frac{L_{\rm H\alpha}}{L_{\rm opt}}
\]

\noindent
where $L_{\rm H\alpha}$ and $L_{\rm opt}$ are the line and continuum luminosities respectively.
From Russell et al. (2006),

\[
L_{\rm OIR} \propto L_{\rm X}^c
\]

\noindent
where the value of $c$ is slightly different for BH and NS systems.
We also fit a crude relation of the form

\[
{\rm EW} \propto L_{\rm X}^{d}
\]

\noindent
above. If we can assume $L_{\rm OIR} \propto L_{\rm opt}$
(i.e. $L_{\rm H\alpha} << L_{\rm opt}$ which is justified for a
maximum EW of $\sim 100$ Angstroms and a filter width of $> 1000$
Angstroms, {\em and} a large fraction of the data in Russell et
al. (2006) being in the R-band) we get

\[
L_{\rm H\alpha} \propto L_{\rm OIR} L_{\rm X}^d
\]

\noindent
Substituting, we get

\[
L_{\rm H\alpha} \propto L_{\rm X}^{c+d} \propto L_{\rm opt}^{1+(d/c)}
\]

\noindent
For the black holes, $c \sim 0.6$ and $d \sim -0.2$, resulting in

\[
L_{\rm H\alpha} \propto L_{\rm X}^{\sim 0.4} \propto L_{\rm opt}^{\sim 0.7}
\]

\noindent
Given the crudeness of the fit to the EW-$L_{\rm X}$ relation, there
seems little point in propagating uncertainties on these relations.
While this analysis is crude, it does demonstrate that the H$\alpha$
line luminosity must vary quite strongly with X-ray luminosity, but not
quite as fast as the continuum.

\section{The origin of the lines and continuum in BHXBs}

A key diagnostic of the emission site of a spectral line is the line
profile itself. A twin-peaked line profile is a strong indication that
the line originates in a rotating flow, such as the atmospheres of
geometrically thin accretion discs (e.g. Horne \& Marsh 1986; note
however that Murray \& Chiang 1996,1997,1998 have shown that accretion
discs with winds can produce single-peaked lines).

In Fig 7 we indicate which lines, from the sample presented in Fig 5,
were reported as twin-peaked. We see that twin-peaked line profiles
have been reported at nearly all luminosities, which strongly suggests
that the line-emitting region is ubiquitously associated with the
rotating accretion flow. As already noted in the introduction, this is
not universally accepted, and a more complex picture is put forward by
Wu et al. (2001, 2002) who argue for three distinct origins for the
emission line depending on the source luminosity and spectral
states. In all cases, the optical continuum and lines are likely to be
excited by irradiation from a central hot continuum source at least at
high luminosities (see arguments in van Paradijs \& McClintock 1994,
and the more recent global study by Russell et al. 2006). As noted in
the introduction, however, it has been suggested that the ionising
source may in fact be (at times) associated with an outflowing jet or
corona and not necessarily a static central X-ray source (see
e.g. Eikenberry et al. 1998; Beloborodov et al. 1999; Markoff et
al. 2005). Although we do not think it has an important impact on this
analysis, we note that double-peaks may be filled in if the
inclination is low or additional components produce low velocity
emission. Similarly, single peaked profiles can also be mainly from a
disc and vice-versa an intrinsic single peaked profiles with central
absorption could give the appearance of a double-peaked line without
needing a disc-like flow.

The observed anti-correlation between H$\alpha$ EW and X-ray luminosity
(both direct and inferred) implies that as the X-ray luminosity
decreases the optical continuum flux drops faster than the line flux.
The analysis in section 2.3 above demonstrates that, over large
luminosity ranges at least, all three quantities (X-ray luminosity,
optical luminosity, H$\alpha$ line flux) are dropping, but at
different rates (X-rays drop fastest, line flux slowest).  The optical
continuum in BHXBs appears to have contributions from both thermal
emission from the outer, irradiated, accretion disc plus, in hard
X-ray states, a component associated with optically thin synchrotron
emission from a jet (Russell et al. 2006). The jet contribution is
strongest at longer wavelengths, dominating in the infrared band, but
probably contributes no more than 30\% of the luminosity in the R-band
(Corbel \& Fender 2002; Homan et al. 2005; Russell et al. 2006), which
contains the H$\alpha$ line.

Note that at low luminosities the companion star may begin to have a
discernible contribution to the continuum luminosity of the
systems. Russell et al. (2006) compile the estimated optical
luminosities of several X-ray binaries, and find that in most systems
the companion only contributes significantly in quiescence
(i.e. $L_{\rm X} \leq 10^{33}$ erg s$^{-1}$). Marsh et al. (1994) show
how in the quiescent system A0620-00 the H$\alpha$ EW is modulated at
the orbital period, presumably due to varying continuum contributions
from the tidally distorted companion.  We do not attempt in this paper
to subtract such a contribution, noting that (i) over the range of
luminosities covered in this compilation it is probably not a major
effect, (ii) any contribution to the continuum by the companion would
serve to {\em increase} the true disc EW at low luminosities and
strengthen the anti-correlation were it taken into account. However, it
is worth noting that in a regime in which an approximately constant
continuum level is set by the companion, the true line flux may be
proportional to the measured EW.

\section{Discussion}

Based on a wealth of observational data, we have an approximate
picture of how the emitted spectra and, to a lesser extent, central
accretion/outflow geometry, might vary in BHXBs, principally as a
function of luminosity.  For example, it is known that an
approximately steady jet is produced at relatively low accretion rates
($\dot{M} / \dot{M}_{\rm Edd} \leq 0.01$) which are associated with
hard X-ray spectra, whereas at higher accretion rates transitions to
softer X-ray states with weaker jets can occur (Fender, Belloni \&
Gallo 2004; Homan \& Belloni 2005). At lower accretion rates a simple
relation between radio and X-ray luminosities appears to hold to at
least $L_X / L_{\rm Edd} \sim 10^{-8}$ (Gallo et al. 2006). There is
however some evidence that the X-ray spectrum does soften at very low
luminosities (Corbel, Tomsick \& Kaaret 2006; Corbel, K\"ording \&
Kaaret 2008). In the case of accreting white dwarfs in cataclysmic
variables, Williams (1980) argued that the outer region of the
accretion discs may at some times be optically thick in the line but
thin in the continuum. If a similar situation exists during the decay
phase of X-ray transients outbursts, it may well contribute
significantly to the observed EW anti-correlation.

The observed anti-correlation is likely to be due to some combination
of these changes in the accretion / outflow geometry, the irradiating
spectrum, and the optical depth in the outer accretion disc. A key
question is whether or not there are changes in the line properties
across the accretion state transitions. As noted earlier this was one
of the key motivations for this research; however, our findings are
inconclusive. On the one hand the line profile and EW of H$\alpha$ in
GX 339-4 is clearly very similar in bright hard or soft states (Fig
1), but the analysis in section 2.1 suggests that the anti-correlation
is better in the hard state below $\sim 0.01 L_{\rm Edd}$. Results for
the black hole transient GRO J1655-40 presented in Shrader et
al. (1996) also suggest a large increase in H$\alpha$ EW across a soft
$\rightarrow$ hard state transition, although the soft X-ray
luminosity of the source would also have been fading during this
period.  The reader is reminded that Wu et al. (2001) claim that the
line profile changes from hard to soft X-ray states.

\begin{figure}
\centerline{\epsfig{file=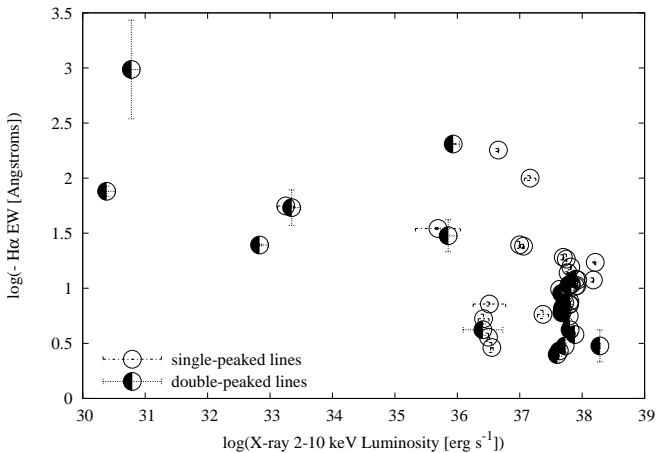, angle=270, width=9cm}}
\caption{Observations in which twin-peaked H$\alpha$ emission has been
resolved. The detection of twin-peaked
emission at all luminosities strongly suggests that the line always
originates in a rotating accretion disc or optically thin accretion flow.}
\end{figure}

\subsection{Optical depth changes}

A possible explanation for the observed anti-correlation is that a
large part of the outer disc becomes cold enough to be optically thin
in the continuum, not the line, during the decline (Williams 1980). In
that case, the optical continuum drops much more quickly than the
ionizing continuum from the inner region (where the inflow is still
optically thick and producing strong continuum emission). As a result
the EW of the optical lines from the outer (optically thin) region
should increase. Such optically thin regions of the outer disc may
well exist in X-ray binaries (e.g. Canizzo \& Wheeler 1984). Shahbaz
et al. (2004) discuss the possibility of regions of different optical
depth contributing to the H$\alpha$ and H$\beta$ emission from the
BHXB A0620-00 in quiescence. It is also possible that there is a
saturation effect, in that during outbursts most of the accretion disc
becomes too hot for the production of H$\alpha$ line emission. The
effect should therefore be different for higher excitation lines such
as HeII.

\subsection{Spectral and geometrical changes}

Although the models mentioned above may well be responsible for the
observed anti-correlation, studies over the past decade have indicated
that the accretion flow in X-ray binaries, in particular black hole
systems, undergo dramatic changes on short timescales which could well
have an effect upon observed emission line profiles. For this reason
we summarize this empirical understanding below, and comment on
whether or not it might affect the observed anti-correlation.

If, as we assume, the H$\alpha$ is a result of irradiation of the
outer disc, then the main source of this irradiation will be EUV
photons with energies between about 13 -- 25 eV. These photons are not
directly observable in nearly all black hole X-ray binaries, which lie
at several kpc in the galactic plane and as a result suffer from
strong interstellar extinction. Lower energy photons cannot ionise
hydrogen, and photons with significantly higher energies (i.e. X-rays)
are more likely to contribute to the reprocessed (reflection)
continuum (e.g. Ross \& Fabian 1993). We do not consider here the
possibility of collisional excitation in e.g. an accretion disc
hotspot.

How plausible is it that changes in the ionising spectrum are
responsible for the observed anti-correlation? In bright soft states
($L_X \geq 0.01 L_{\rm Edd}$) the X-ray spectra of BHXBs are dominated
by thermal (accretion disc) components with kT $\geq 1$ keV. In less
bright, but still very luminous ($10^{-4} L_{\rm Edd} \leq L_X \leq
10^{-2} L_{\rm Edd}$), hard X-ray states the X-ray spectrum is
dominated by a component which peaks at $\sim 100$ keV and probably has
its origin in a Comptonising `corona' (e.g. Sunyaev \& Titarchuk
1980), with some possible contributions from a jet (e.g. Markoff,
Nowak \& Wilms 2005).  It should be noted that there can be strong
hysteresis between spectral state and luminosity for $L_X \ga 10^{-2}
L_{\rm Edd}$ (Homan \& Belloni 2005).  As a source drops further in
luminosity the accretion disc temperature should drop monotonically
until in `quiescence' ($L_X \sim 10^{-8.5} L_{\rm Edd}$) it is only
about 1 eV (McClintock, Horne \& Remillard 1995; McClintock et
al. 2003). As noted above, at most X-ray luminosities this thermal
disc component is essentially unmeasurable as it peaks at (E)UV
wavelengths, in which band most distant galactic plane BHXBs cannot be
observed. In the case of XTE J1118+480 (McClintock et al. 2003) this
disc component was found to be significantly more luminous than the
X-ray component when in quiescence. Therefore, at the lowest X-ray
luminosities the bulk of the photons released by the accretion process
are not able to ionise hydrogen, which -- naively -- should result in
a reduced H$\alpha$ EW at the lowest luminosities, {\em contrary} to
what is observed. However, the optical continuum is also dominated by
reprocessing and so the overall effect on EW will depend upon which
component is pumped most effectively by the ionizing continuum, which
in some sense returns us to the models dealing with the optical depth
of the outer disc (e.g. Williams 1980).

What about changes in the geometry of the accretion flow and outflow
with luminosity? Some recent sketches of the accretion flow as a
function of luminosity for BHXBs (Esin, McClintock \& Narayan 1997;
Done, Gierlinski \& Kubota 2007) suggest that as luminosity increases
the ability of the central hard X-ray source to illuminate the disc is
reduced, as the coronal component shrinks.  Geometries which include
jets (e.g. Fender, Belloni \& Gallo 2004; Ferreira et al. 2006) may be
more complete, although it is unclear how significantly any X-ray
emission from the jet could contribute to irradiation of the accretion
disc or its atmosphere (Markoff, Nowak \& Wilms 2005). Such X-ray
emitting regions near the base of the jet may be essentially
indistinguishable from an outflowing corona (e.g. Beloborodov
1999). Nevertheless, it has been suggested that the velocity of the
outflow from X-ray binaries may increase with luminosity, and in a
more complex way with spectral state. If the jet or outflowing corona
is responsible for some irradiation of the disc, then an increase in
velocity with luminosity could result in a reduced H$\alpha$ EW as an
increasing fraction of the X-rays are beamed along the direction of
motion, away from the disc.

\section{Conclusions}

A prime motivation for this research was to investigate whether the
profiles of optical emission lines change significantly across the
accretion state transitions in X-ray binary systems. Such a result may
have indicated a strong response to a changing irradiating spectrum
and/or geometry of the line-emitting region. However, the results in
this respect remain inconclusive.

What we did find, and report in detail, is an anti-correlation between
the broadband luminosity (optical -- X-ray) and the equivalent width
of the H$\alpha$ emission line in X-ray binaries. Possibly the most
closely related phenomenon already known is a comparable
anti-correlation in the H$\beta$ line for Cataclysmic Variables
(Patterson 1984; see also Witham et al. 2006), accreting binaries
hosting a white dwarf rather than a black hole or neutron
star. Patterson (1984) compares this result with the models of Tylenda
(1981), which are similar to those of Williams (1980) in which the
outer accretion disc becomes optically thin at low luminosities.  This
may also turn out to be the appropriate explanation for the X-ray
binaries under discussion here, but the complex and hysteretical
accretion state changes about which we have learned much in the past
decade caution us against drawing such a conclusion at this time.
More detailed observations of line profiles across state transitions
would clearly be of interest. The reader is further reminded that (i)
rebrightenings during the decay of transients, (ii) the monotonic
behaviour of QPOs in sources such as XTE J1118+480 despite varying
behaviour in $L_X$, and (iii) the poorer anticorrelation of EW with
quasi-simultaneous $L_X$ than with time all suggest that the 
causal link between luminosity and EW may be complex. 

It it worth mentioning in passing the similarity with the Baldwin
Effect observed in accretion flows around supermassive black holes
(Baldwin 1977; see also e.g. Mushotzky \& Ferland 1984). In this case
an anti-correlation is observed between the EW some lines originating
in the broad line region (BLR) and the continuum luminosity. A related
effect is observed in the X-ray band (Iwasawa \& Taniguchi 1993; Page
et al. 2004), and maybe also in the strong stellar winds of Wolf-Rayet
stars (Morris et al. 1993). However, there is little evidence for a
BLR-like region in X-ray binary systems, probably due to the high
ionisation state of the gas in the inner disc region (e.g. Proga,
Kallman \& Stone 2000). Therefore the physical origins of the two
effects are almost certainly rather different.

Further observations would obviously be of interest to understand
better this effect. These should include observations of other lines
and in other bands. Observations in the near-infrared, where the
continuum should have a much stronger jet contribution (e.g. Corbel \&
Fender 2002; Russell et al. 2006) should show a much stronger reaction
to accretion state changes (e.g. we would expect a large jump in EW
of a line like Br $\gamma$ as the jet switches off in a hard
$\rightarrow$ soft state transition). It is interesting to note that
the spectra in Fig 1 hint at a similar anti-correlation for HeI, and
HeII would be even more interesting, given the higher ionisation
potentials.  The observed anti-correlation holds the promise of being
able to estimate the luminosity of a source from the H$\alpha$ EW,
something which could prove invaluable in distance determinations for
faint X-ray binaries. However, the relation obviously needs to be
significantly improved before that is possible, and may turn out to
have enough intrinsic scatter to reduce its usefulness in this aspect
(as is in fact the case for the Baldwin Effect in AGN). In any case it
is promising that low-luminosity accreting sources should be clearly
identifiable in e.g. H$\alpha$ surveys by their large equivalent
widths.

Finally, it is interesting to note that observations of the variation
of the H$\alpha$ line strength with luminosity may in fact be our best
way to track the behaviour of the accretion disc component at
intermediate luminosities. Below $L_X \sim 10^{-3} L_{\rm Edd}$ the
accretion disc component peaks in the (E)UV spectral regime and is
unobservable for nearly all sources.  For two sources only has it been
observed in quiescence, at which point it appears that it no longer
extends to the ISCO but is truncated (McClintock et al. 1995,
2003). Although well-defined models exist (e.g. Dubus, Hameury \&
Lasota 2001), exactly how and where this truncation begins is
observationally very uncertain (e.g. Miller et al. 2006). However,
since the H$\alpha$ line responds preferentially to photons of
energies 13--25 eV, it may be the best way to study the evolution of
this disc component at luminosities $10^{-6} \la L_X / L_{\rm Edd} \la
10^{-3}$.

\section*{Acknowledgements}

We would like to thank Cristina Zurita for providing tables of data
for XTE J1118+480, and Tariq Shahbaz and Gavin Dalton for providing
some of the data for Fig 1.  We would like to thank Guillaume Dubus,
Daniel Proga and Kinwah Wu for careful reading of, and useful comments
on, an earlier version of this work, and the anonymous referee for
advice which helped to improve this work.

\end{document}